\documentclass[pra,twocolumn,aps]{revtex4}
\usepackage{amsmath}
\usepackage{amssymb}
\usepackage{graphicx}

\begin{document}
\title{Local field-interaction approach to the Dirac monopole 
}
\author{Kicheon Kang}
\email{kicheon.kang@gmail.com}
\affiliation{Department of Physics, Chonnam National University, Gwangju 61186, 
 Republic of Korea}

\begin{abstract}
We introduce the local field interaction approach to Dirac magnetic monopoles.
Our analysis reveals two physically different types of a monopole.
The first type is free of singularity, and
the field angular momentum plays an essential role in the interaction. 
The second type 
is described as an endpoint of an invisible semi-infinite flux tube
(a Dirac string). 
Notably, a different phase factor $(-1)^n$ exists between  
the two types where $n$ is the quantum number of the field angular
momentum. Our study 
provides a realistic description of the two types of monopoles.
Various aspects of these monopoles are discussed,
including the Maxwell dual of the Dirac string, exchange symmetry, and an
analogy to the Coriolis interaction. 
\end{abstract}

\maketitle

\section{Introduction}
Dirac~\cite{dirac31} originally showed that the existence 
of a magnetic monopole is consistent with quantum theory 
if the magnetic charge $g$ satisfies the quantization condition 
\begin{equation}
 eg/(2\hbar c) = n ,
\label{eq:dirac}
\end{equation}
where $e$ is the electric charge of another particle and $n$ is an integer.
The quantum mechanical description of a charge under the vector 
potential ${\mathbf A}$
generated by the monopole inevitably includes a singularity 
(or discontinuity) in ${\mathbf A}$ because
$\nabla\cdot(\nabla\times{\mathbf A}) \neq 0$.
Eq.~\eqref{eq:dirac} is derived using the 
single-valuedness of the wave function. 
Recently, by adopting a local field interaction~(LFI) approach, 
we developed a quantum theory of electromagnetic interaction
that does not involve ${\mathbf A}$~\cite{kang13,kang15}. 
Classical electrodynamics and the topological
Aharonov-Bohm effect are successfully reproduced in the LFI theory. 
In addition, the remarkable consequences of the LFI theory concerning the 
locality~\cite{kang17,kim18} and
the gauge symmetry~\cite{kang19} were also revealed. 
In the LFI approach, the role of the potential is replaced 
by the field momentum produced by the charge and the external magnetic field
$\mathbf B$. 

In this work, we apply the LFI theory to the problem of a charge 
interacting with
a magnetic monopole and show that two types of monopole, 
namely Type I and Type II, are possible~(see Fig.~1).
It should be noted that the difference is not merely a mathematical artifact 
but a physical reality.
The LFI approach involves replacing the vector potential by the field momentum.
In the case of a charge-monopole pair, however, the field momentum vanishes and
thus fails to describe the interaction between the two particles.
We show that the interaction is mediated by the 
field angular momentum produced by two particles (electric charge
and magnetic monopole). 
The field angular momentum plays the same role as the spin
and is essential for constructing the singularity-free description 
of the monopole (``Type I").
Alternatively, it is also possible to describe the monopole with
a Dirac string in the LFI approach (``Type II").
In monopoles of the latter type, 
the interaction between the two particles is produced by the
field momentum confined inside the string. 
The Type-II monopole is equivalent to the original description of the
Dirac monopole, which has a singular string.
The two different types of monopole reveal duality in the classical
and quantum equations of motion. 
Notably, the two types are not completely equivalent: 
an additional phase factor $(-1)^n$ appears
in the quantum state of Type I. The implication thereof is discussed in detail.

Note that we do not consider 
singularity-free monopoles in the context of
non-Abelian gauge group and spontaneous symmetry 
breaking~\cite{thooft74,polyakov74}. 
Our discussion is restricted to the original Dirac monopole.

The paper is organized as follows.
In Sec.~II, we derive the Lagrangian for a charge-monopole pair 
using the LFI approach.
We point out that two different types of monopole 
are possible, depending on whether a Dirac string is present. 
Sec.~III presents the evaluation of the quantum mechanical phase shifts 
for each type.
The field angular momentum and the singularity play major roles 
in the different phase factors of the two types.
In Sec.~IV, the corresponding Hamiltonians are derived
for each of the two types and their duality is analyzed.
Sec.~V discusses a few intriguing aspects 
derived from our approach.
A notable property of the Type-II monopole is found from the Maxwell
duality: 
the electric charge can also be described as an endpoint of 
singularity. 
The exchange symmetry of the charge-monopole composite particles is
derived for the two types.
In addition, the formal equivalence between the Type-I monopole and the
Coriolis interaction is demonstrated. Sec.~VI concludes the paper.

\section{Field-interaction Lagrangian of a charge-monopole pair}
In the standard potential-based approach, 
the dynamics of a charge ($e$) under an external magnetic field~(${\mathbf B}$) 
is described by the Lagrangian 
\begin{equation}
 {\cal L}_A = {\cal L}_0 + \frac{e}{c} \dot{\mathbf{r}}\cdot\mathbf{A} \,,
\label{eq:L_A}
\end{equation}
where ${\cal L}_0 = -mc\sqrt{c^2-\dot{\mathbf{r}}\cdot\dot{\mathbf{r}}}$ 
is the kinetic part of the charged particle with mass $m$. 
The vector potential $\mathbf{A}$ in the interaction term 
is replaced by the field momentum
($\mathbf{\Pi}$) in the LFI theory, 
and the Lagrangian of the system is given by~\cite{kang13,kang15}
\begin{subequations}
\label{eq:L_Pi}
\begin{equation}
 {\cal L}' = {\cal L}_0 + \dot{\mathbf{r}}\cdot \mathbf{\Pi} 
\end{equation}
where the field momentum, 
\begin{equation}
 \mathbf{\Pi} = \frac{1}{4\pi c}\int \mathbf{E}_e\times\mathbf{B} 
 d^3\mathbf{x}, 
\label{eq:Pi}
\end{equation}
\end{subequations}
is generated by the overlap between the electric field ($\mathbf{E}_e$) 
of charge $e$ and the external $\mathbf{B}$. 
In the absence of the Dirac-string-type singularity~(``Type I" in Fig.~1(a)), 
$\mathbf{\Pi}=0$ for the charge-monopole pair 
(see {\em e.g.}, Section 6.12 of Ref.~\onlinecite{jackson99}), 
in which case the Lagrangian \eqref{eq:L_Pi} fails to describe 
the charge-monopole interaction.

The appropriate LFI Lagrangian for the charge-monopole
pair is achieved by including the rotational degree of freedom.
For the angular velocity $\dot{\vec{\psi}}$ of the charge, we introduce the
Lagrangian
\begin{subequations}
 \label{eq:L_psi}
\begin{equation}
 {\cal L} = {\cal L}_0 + \dot{\vec{\psi}} \cdot \mathbf{S} , 
\end{equation}
where $\mathbf{S}$ is the electromagnetic field angular momentum 
produced by $e$ and $g$.
(see e.g., Ref.~\onlinecite{jackson99}):
\begin{equation}
 \mathbf{S} = -\frac{eg}{c} \hat{\mathbf{r}} .
\label{eq:Lem-monopole}
\end{equation}
\end{subequations}
Here, $\mathbf{S}$ plays the same role
as the particle spin, as is discussed later. 
A microscopic derivation of this Lagrangian is presented below in this
section.
First, the validity of Eq.~\eqref{eq:L_psi} is verified by deriving
the classical equation of motion.
The Lagrange equation for the angle variable $\vec{\psi}$, 
\begin{equation}
 \frac{d}{dt}\left( \frac{\partial \cal L}{\partial\dot{\vec{\psi}}} \right) 
 - \frac{\partial \cal L}{\partial\vec{\psi}} = 0 \,,
\end{equation}
leads to the equation of motion
\begin{subequations}
\label{eq:eom}
\begin{equation}
 \frac{d}{dt} \left( \mathbf{L}+\mathbf{S} \right) = 0 \,,
\end{equation}
where 
$\mathbf{L}=\partial{\cal L}_0 / \partial \vec{\psi} $ is the
kinetic angular momentum of the charge.
This indicates the conservation of the total angular momentum. 
It can be rewritten in a more
familiar form involving the Lorentz force, as 
\begin{equation}
 \frac{d\mathbf p}{dt} = \frac{e}{c} \dot{\mathbf{r}} \times \mathbf{B} \,,
\label{eq:eom_Lorentz}
\end{equation}
\end{subequations}
where $\mathbf{p} \equiv \partial {\cal L}_0/\partial\dot{\mathbf r}$ is
the kinetic momentum of the charge, and
$\mathbf{B} = g\hat{\mathbf r}/r^2$ is the magnetic field generated by the
monopole.

The above derivation of the equation of motion \eqref{eq:eom}
demonstrates the validity of the Lagrangian \eqref{eq:L_psi} 
in describing the charge interacting with the magnetic
monopole. Now, we present a microscopic derivation of
the Lagrangian \eqref{eq:L_psi} based on the LFI 
approach~\cite{kang13,kang15}.
For simplicity, we do not consider the motion of the monopole at this stage. 
(The Lagrangian including the motion of the monopole is discussed in
Sec.~V-A).
In the LFI approach, a charged particle subject to an 
external electromagnetic field
is described by the Lagrangian:
\begin{subequations}
\begin{equation}
 {\cal L} = {\cal L}_0 + {\cal L}_\mathrm{in} ,
\end{equation}
where
\begin{equation}
 {\cal L}_\mathrm{in} = \frac{1}{8\pi} 
    \int F_{\mu\nu}^{(e)} F^{\mu\nu} d^3\mathbf{x},
\label{eq:L_in}
\end{equation}
\end{subequations}
represents the interaction between the field of charge and the external field
(denoted by the field tensors $F_{\mu\nu}^{(e)}$
and $F^{\mu\nu}$, respectively).
In our case, $F^{\mu\nu}$ is produced by the monopole.

For a moving charged particle with velocity $\dot{\mathbf r}$, the interaction
part of the Lagrangian \eqref{eq:L_in} can be expressed as 
${\cal L}_\mathrm{in} = \dot{\mathbf r}\cdot\mathbf{\Pi}$~\cite{kang13,kang15}. 
However, as mentioned above, $\mathbf{\Pi} =0$ in our system.
That is, the translational motion does not produce
a coupling with the monopole.  Instead, a nonvanishing 
field interaction is generated in the rotational degree of freedom.
The rotation of charge, with its angular velocity $\dot{\vec{\psi}}$, produces
a magnetic field $\mathbf{B}_e(\mathbf{x})$ at a position 
$\mathbf{x}$~(see Fig.~2) in the form
\begin{displaymath}
 \mathbf{B}_e(\mathbf{x}) = \frac{1}{c} 
     (\dot{\vec{\psi}}\times\mathbf{x}) \times \mathbf{E}_e(\mathbf{x}) ,
\end{displaymath}
where $\mathbf{E}_e(\mathbf{x})$ is the electric field of the charge.
This expression of $\mathbf{B}_e$ is obtained by evaluating 
the magnetic field in a uniformly rotating frame with angular velocity 
$-\dot{\vec{\psi}}$.
The nonvanishing term in the interaction Lagrangian~\eqref{eq:L_in} originates
from the $\mathbf{B}_e\cdot\mathbf{B}$ coupling, and we obtain 
\begin{subequations}
\begin{equation}
 {\cal L}_\mathrm{in} = \dot{\vec{\psi}} \cdot \mathbf{S} \,,
\end{equation} 
where 
\begin{equation}
 \mathbf{S} = \frac{1}{4\pi c} \int \mathbf{x} 
    \times \left( \mathbf{E}_e\times \mathbf{B} \right) d^3\mathbf{x} 
\end{equation}
\end{subequations}
corresponds to the field angular momentum. $\mathbf{S}$ is reduced to 
Eq.~\eqref{eq:Lem-monopole} for the magnetic field of the monopole. 
%
The Lagrangian~\eqref{eq:L_psi} derived above does not include any
singularity and is classified as ``Type I". 

It is also possible to describe the monopole attached to a Dirac string 
(``Type II")
in the LFI approach with the Lagrangian \eqref{eq:L_Pi}.
The string is an invisible tube of the magnetic flux of $4\pi g$
terminated at the origin~(Fig.~1(b)). For the string located at $\theta=\pi$,
an evaluation of Eq.~\eqref{eq:Pi} shows that the field momentum 
\begin{equation}
 \mathbf{\Pi}(\mathbf{r}) 
    = \frac{eg}{c} \frac{1-\cos{\theta}}{r\sin{\theta}}\hat{\phi} 
\label{eq:Pi_string}
\end{equation} 
is generated inside the string.
The classical equation of motion \eqref{eq:eom_Lorentz} is derived from 
the Lagrangian \eqref{eq:L_Pi} with $\mathbf{\Pi}$ of Eq.~\eqref{eq:Pi_string},
demonstrating the duality of the two
different types represented by the Lagrangians \eqref{eq:L_Pi} and
\eqref{eq:L_psi}, respectively.
The Type-II monopole described by the Lagrangian \eqref{eq:L_Pi}
is equivalent to the original problem of the Dirac monopole with singular
vector potential.

The location of the Dirac string gives rise to a problem of arbitrariness. 
The value of the field momentum depends on the position of the string.
However, two field momenta $\mathbf{\Pi}'$ and $\mathbf{\Pi}$ with
different locations of the string are related
by the gauge transformation
$\mathbf{\Pi}' = \mathbf{\Pi} + \nabla\chi$ with a single-valued scalar
function $\chi$. This property of the gauge transformation is also present
in the vector potential of the ordinary formulation~(see {\em e.g.}, 
Section 6.12 of Ref.~\onlinecite{jackson99}).
In any case, the physical observables are independent of the position of
the string.

\section{Quantum theory of monopoles with 
two types of the field-interaction Lagrangian}
%
Here, we apply the Lagrangians derived in Sec.~II~(Eqs.~\eqref{eq:L_Pi} and
\eqref{eq:L_psi}) to quantum theory.
We consider an arbitrary closed loop in the path of the charge (Fig.~3). 
In Type I~(Fig.~3(a)), the phase shift~($\varphi$) generated by the monopole 
is evaluated from the interaction term of the Lagrangian~\eqref{eq:L_psi}: 
\begin{equation}
 \varphi = \frac{1}{\hbar}\oint \dot{\vec{\psi}}\cdot \mathbf{S} dt 
         = \frac{1}{\hbar}\oint \mathbf{S} \cdot d\vec{\psi} \,.
\end{equation}
For an arbitrary loop with a solid angle $\Omega$, we find
\begin{equation}
 \varphi = \frac{S}{\hbar} (\Omega-2\pi) .
\end{equation}
The phase shift can also be evaluated from the opposite side of the solid 
angle, $\bar{\Omega}=-(4\pi-\Omega)$: 
$ \bar{\varphi} = S(\bar{\Omega}-2\pi)/\hbar$.
The equivalence of the two phases, with modulo $2\pi$ 
($\bar{\varphi}=\varphi+2n\pi$ with integer $n$), 
imposes the quantization of the field angular momentum
\begin{equation}
 S = eg/c = n\hbar/2 \,.
\label{eq:L_em}
\end{equation}
This is exactly the Dirac quantization for the electric and magnetic
charges. With this quantization, the phase shift is given by
\begin{equation}
 \varphi = \frac{eg}{\hbar c}\Omega - n\pi .
\label{eq:varphi}
\end{equation}

Notably, we find a different phase shift ($\varphi'$)
from the Type-II Lagrangian \eqref{eq:L_Pi},
\begin{equation}
 \varphi' = \frac{1}{\hbar} \oint \mathbf{\Pi}\cdot d\mathbf{r}
   = \frac{eg}{\hbar c} \Omega \,,
\label{eq:varphi'}
\end{equation}
which corresponds to the geometric phase, or the 
Aharonov-Bohm phase generated by the magnetic flux of the monopole.
The phase shift $\varphi'$ can also be obtained from the usual 
potential-based Lagrangian \eqref{eq:L_A}.
The additional phase shift $-n\pi$ in Type I
(Eq.~\eqref{eq:varphi})
reflects the fermionic (bosonic) nature of the
field angular momentum for odd (even) values of $n$ (Eq.~\eqref{eq:L_em}).
For odd values of $n$, this gives rise to an additional phase factor of $-1$.
This phase factor originates from the field angular momentum
and is unrelated to the intrinsic spin of each particle.

The difference between the two phases, $\varphi$ (Eq.~\eqref{eq:varphi}) 
and $\varphi'$ (Eq.~\eqref{eq:varphi'}), 
is also closely related to the absence or presence of a Dirac string. 
The Type-II monopole is not an isolated particle but 
emerges as an endpoint of the string. 
This implies that
another monopole with opposite magnetic charge $-g$ exists 
somewhere at a large distance from the system.
This gives rise to additional field angular momentum of $n\hbar/2$ and the
phase shift of $n\pi$, which cancels $-n\pi$ in Eq.~\eqref{eq:varphi}.
This cancellation cannot be avoided, even when the string stretches
to infinity, and explains the difference between 
$\varphi$(Eq.~\eqref{eq:varphi}) and $\varphi'$ (Eq.~\eqref{eq:varphi'}).

Our analysis based on the LFI approach clearly
suggests that there could be two different types of magnetic monopole: 
(i) ``Type I" without singularity, which can be described in terms
of the interaction between the ``spin" of the charged particle and the
field angular momentum (Eq.~\eqref{eq:L_psi}); 
(ii) ``Type II" with a singular Dirac string
where the motion of the charge couples to the field momentum
localized in the string
(Eqs.~\eqref{eq:L_Pi} and \eqref{eq:Pi_string}). 
Remarkably, this classification is not merely a mathematical
construction for describing the same system. For odd multiples of the
field angular momentum, the two types can be 
distinguished by the presence (absence) of the additional phase shift
$n\pi$ in $\varphi$ ($\varphi'$) of Type I (Type II).

\section{Hamiltonian and the duality of the two types of monopoles}
\subsection{Hamiltonian}
Derivation of the Hamiltonian from the
Lagrangian of the system via a Legendre transformation is straightforward.
For Type I~(Eq.~\eqref{eq:L_psi}), 
four independent variables are necessary,
$q_i = (r,\vec{\psi})$, composed of the distance from the origin ($r$) and
the three-dimensional angle vector $\vec{\psi}$. 
The Hamiltonian is obtained from
the relation $H = \sum_i \dot{q}_i p_i - \cal L$, where 
$p_i = \partial{\cal L}/\partial\dot{q}_i$ is the conjugate momentum.
We find
\begin{equation} 
 H = \sqrt{c^2\left( p_r^2 + 
      \frac{(\mathbf{J}-\mathbf{S})^2}{r^2} 
              \right) + m^2c^4 } ,
\label{eq:H}
\end{equation}
where $p_r=\partial{\cal L}/\partial\dot{r}$ and
$\mathbf{J} = \partial{\cal L}/\partial\dot{\vec{\psi}}$
denote the radial component of the canonical momentum and 
the angular momentum vector, respectively. 
When applied to quantum theory,
$\mathbf{J} = (\hbar/i)\partial/\partial\dot{\vec{\psi}}$, 
and $\mathbf{S}$ becomes
a spinor satisfying the quantization condition of Eq.~\eqref{eq:L_em}.
In the nonrelativistic limit, the Hamiltonian is reduced to
\begin{equation}
 H = mc^2 
      + \frac{1}{2m} \left( p_r^2 + 
        \frac{(\mathbf{J}-\mathbf{S})^2}{r^2} 
                     \right) \,. 
\label{eq:H_nonrel}
\end{equation}
In the present case, $\mathbf{S}$ is parallel to $\mathbf{r}$ and satisfies
$(\mathbf{J}-\mathbf{S})^2 = \mathbf{J}^2-\mathbf{S}^2$. 
Under this condition, the Hamiltonian
\eqref{eq:H_nonrel} is equivalent to that adopted in the
spin approach to the monopole in Ref.~\onlinecite{goldhaber65}.
The ``spin Hamiltonian" in Ref.~\onlinecite{goldhaber65} 
was introduced for consistency in the classical equation of motion.
In contrast to this, we presented
a microscopic derivation of the Hamiltonian 
(\eqref{eq:H} and \eqref{eq:H_nonrel}) above.
The Type-II Hamiltonian, denoted by $H'$, 
can also be obtained from the Lagrangian~\eqref{eq:L_Pi}
as
\begin{equation}
 H' = \sqrt{c^2 (\mathbf{p}-\mathbf{\Pi})^2 + m^2c^4 } \,,
\label{eq:H'}
\end{equation}
where $\mathbf{p}=\partial{\cal L}/\partial\mathbf{r}$ is the canonical
momentum.
This Hamiltonian contains a singularity (Dirac string) in $\mathbf{\Pi}$ 
(see e.g., Eq.~\eqref{eq:Pi_string}).

\subsection{Duality}
The duality of the two types of monopoles is already apparent
in the Lagrangian approach in the previous sections.
The two Lagrangians representing each type (Eqs.~\eqref{eq:L_psi} and
\eqref{eq:L_Pi}) provide the same classical dynamics and the quantum
phase shift for an arbitrary closed path except the additional 
$-n\pi$~(Eq.~\eqref{eq:varphi}) in Type-I monopole. 
In the following, the duality of the two types is derived in a general way 
from the Hamiltonians $H$~(Eq.~\eqref{eq:H}) and 
$H'$~(Eq.~\eqref{eq:H'}).

Let $ u_{i\rightarrow f} \equiv 
    \langle \mathbf{r}_f, t_f|\mathbf{r}_i, t_i \rangle $ 
be the transition amplitude from an initial
($\mathbf{r}_i, t_i$) to the final ($\mathbf{r}_f, t_f$) spacetime locations 
for a Type-I system.
In the Feynman path-integral representation, it reads
\begin{equation}
 u_{i\rightarrow f} = \int_{\mathbf{r}_i}^{\mathbf{r}_f}{\cal D}[\mathbf{r}(t)]
    e^{ (i/ \hbar) \int{\cal L}\,dt } \,, 
\label{eq:u_if}
\end{equation}
where ${\cal L}$ is the Type-I Lagrangian \eqref{eq:L_psi}.
The same transition amplitude, namely, $u'_{i\rightarrow f}$,
can be defined for a Type-II system 
where ${\cal L}$ in Eq.~\eqref{eq:u_if} is replaced
by ${\cal L}'$ of the Lagrangian \eqref{eq:L_Pi}.

Let $u_1$($u'_1$) be the transition amplitude in the Type-I (Type-II) system
for a closed path, that is,
 $u_1=u_{i\rightarrow f}$ ($u'_1=u'_{i\rightarrow f}$) 
for $\mathbf{r}_f=\mathbf{r}_i$ with one-loop rotation. 
We find 
\begin{equation}
  u'_1 = (-1)^n u_1 \,,
\label{eq:u1}
\end{equation}
from the relation between $\varphi$ (Eq.~\eqref{eq:varphi}) and 
$\varphi'$ (Eq.~\eqref{eq:varphi'}). The transition amplitude of 
Eq.~\eqref{eq:u_if} is also expressed in the Hamiltonian representation as
\begin{eqnarray}
 u_{i\rightarrow f} &=& \langle \mathbf{r}_f | e^{-iH(t_f-t_i)/\hbar} |
      \mathbf{r}_i \rangle  \nonumber \\
  &=& \sum_l \psi(\mathbf{r}_f)\psi^*(\mathbf{r}_i)\, e^{-iE_l(t_f-t_i)/\hbar} 
    \,,
\end{eqnarray}
where $E_l$ and $\psi_l$ denote the eigenvalue and eigenfunction
of $H$, respectively.
We find that the two types are related 
by a unitary transformation ($U$) 
\begin{equation}
\psi \rightarrow \psi' = U\psi \;\;\; \mbox{\rm along with} \;\;\; 
H \rightarrow H' = UHU^\dagger \,,
\end{equation}
and the transition amplitude in 
Type-II representation can be expressed as
\begin{equation}
 u'_{i\rightarrow f} = \sum_l U^\dagger(\mathbf{r}_i) U(\mathbf{r}_f) 
    \psi(\mathbf{r}_f)\psi^*(\mathbf{r}_i)\, e^{-iE_l(t_f-t_i)/\hbar}
    \,.
\end{equation}

The general condition of the unitary transformation $U$ can be imposed from
Eq.~\eqref{eq:u1}. By writing 
\begin{subequations}
\begin{equation}
 U(\mathbf{r}) = e^{-i\int^\mathbf{r} \mathbf{a}\cdot d\mathbf{r}'} \,,
\end{equation}
we obtain
\begin{equation}
 \oint \mathbf{a}\cdot d\mathbf{r} = n\pi \,.
\end{equation}
from Eq.~\eqref{eq:u1}. 
For example, the unitary operation
\begin{equation}
 U = e^{-i\mathbf{S}\cdot\vec{\phi}/\hbar}
\end{equation}
\end{subequations}
transforms $H$ (Type I) into $H'$ (Type II) (see also 
Ref.~\onlinecite{goldhaber65}).

The analysis in Sections III and IV clearly indicates that 
both types of monopoles are consistent with quantum theory under the same
Dirac quantization condition.
We cannot predict the type of the real monopoles.
On the other hand, for effective monopoles, their types can be classified 
in our scheme. 
It is found that a spin (pseudospin) system behaves similar to
a charged particle under the magnetic field of a monopole,
as manifested in Berry's phase~\cite{berry84} 
(see Refs.~\onlinecite{shapere89,xiao10} for a review).
This case belongs to Type I, where its spin (pseudospin) is
represented by the spinor in the Hamiltonian (Eq.~\eqref{eq:H}). 
The Type-II system includes effective monopoles produced by analogues of
the Dirac string: an endpoint of a long solenoid or others of a similar
nature.
Examples can be found in various systems such as 
spin ice~\cite{castelnovo08,morris09,bramwell09} and
synthetic magnetic field~\cite{ray14}, {\em etc.}. 

\section{Discussion}
\subsection{Various configurations of the singularity}
As described above, a notable difference exists
between the two types of monopole description: the absence (presence)
of a singularity in the form of a Dirac string in Type I (Type II). 
In either case, the
singularity is unobservable. Nevertheless, we can gain insight 
into this problem 
by considering the dynamics of both particles on an equal footing and
exchanging the role of electric ($e$) and magnetic ($g$) charges, 
especially in Type II.

The symmetry of the electrodynamics under Maxwell duality transformation
(see e.g., Ref.~\onlinecite{dowling99}) leads to an intriguing consequence
on the Type-II monopole as described below.
In the Maxwell dual (Fig.~4(b)) of the original Type-II system (Fig.~4(a)), 
the ``Dirac
string" is attached to the electric charge. In other words, electric
charge emerges as an endpoint of an invisible string of the electric flux. 
The field momentum $\mathbf{\Pi}$ in the dual configuration is
generated by the overlap between the electric string
and the magnetic field generated by the monopole. In both cases of Fig.~4(a,b), 
the Lagrangian of the system is expressed in the form
\begin{equation}
 {\cal L}' = {\cal L}_0 + (\dot{\mathbf{r}} - \dot{\mathbf{r}}_g) 
            \cdot \mathbf{\Pi} \,,
\end{equation}
where ${\cal L}_0$ includes the kinetic part of both the particles, and 
$\mathbf{r}_g$ is the velocity of the monopole $g$.
The field momentum depends on
the location of the strings. For the particular case in which the string is
located at $\theta=\pi$, the momentum is expressed by Eq.~\eqref{eq:Pi_string} 
in both the original (Fig.~4(a)) and dual configurations (Fig.~4(b)).
Moreover, it is also possible that both particles are the endpoints
of the corresponding strings (Fig.~4(c)).
In this case, the Lagrangian is given by 
\begin{equation}
 {\cal L}' = {\cal L}_0 + \dot{\mathbf{r}}\cdot\mathbf{\Pi}_g 
           - \dot{\mathbf{r}}_g \cdot \mathbf{\Pi}_e \,,
\end{equation}
involving two different field momenta, $\mathbf{\Pi}_g$ and $\mathbf{\Pi}_e$, 
which are localized inside the magnetic and electric Dirac strings, 
respectively. Irrespective of the 
configuration, the physics of all cases (Fig.~4(a,b,c)) is equivalent, 
leading to the
same classical equation of motion and the quantum phase shift 
(Eq.~\eqref{eq:varphi'}).

The problem is much simpler in Type I because the Maxwell dual is identical
to the original system itself.
The dynamics of the two
particles can be described on an equal footing by generalizing
the Lagrangian \eqref{eq:L_psi} as
\begin{equation}
 {\cal L} = {\cal L}_0 
    + (\dot{\vec{\psi}}-\dot{\vec{\psi}}_g)\cdot\mathbf{S} \,, 
\end{equation}
where $\dot{\vec{\psi}}_g$ represents the angular velocity of the monopole.
Apparently, the system is identical upon the exchange of $e$ and $g$ 
in the absence of a Dirac string.

\subsection{Exchange symmetry of charge-monopole composites}
Let us consider two identical charge-monopole composite ``particles" 
with each one
located at $\mathbf{r}_1$ and $\mathbf{r}_2$. The wave function of the
system may be written as $\Psi=\Psi(\mathbf{r},\eta_1,\eta_2)$,
where $\mathbf{r}\equiv \mathbf{r}_1-\mathbf{r}_2$ 
and $\eta_i$ ($i=1,2$) represents the
internal state of each particle.
For the exchange of the two particles (represented by
the operator $P$), we show that 
\begin{equation}
 P\Psi(\mathbf{r},\eta_1,\eta_2) \equiv \Psi(-\mathbf{r},\eta_2,\eta_1) 
  = (-1)^{2s+n} \Psi(\mathbf{r},\eta_1,\eta_2) \,
\label{eq:exchange}
\end{equation}
where $s$ and $n$ are the intrinsic spin and the Dirac quantum number
(Eq.~\eqref{eq:dirac}) of the composite, respectively.
We find that this result is independent of the types of the
monopoles. Interestingly, we arrive at the result (Eq.~\eqref{eq:exchange})
in different ways for Type-I and Type-II monopoles as is shown below.

In general, the two Lagrangians ${\cal L}$ and ${\cal L}_0$ with the relation
\begin{subequations}
\label{eq:L_transform}
\begin{equation}
 {\cal L} = {\cal L}_0 + \frac{d\Lambda}{dt} 
\end{equation}
exhibit a particular symmetry.
With the aid of the Feynman path-integral formulation, the transition
amplitudes $u$ and $u_0$ (associated with the Lagrangians ${\cal L}$ and
${\cal L}_0$, respectively,) have the relation
\begin{equation}
 u = u_0 e^{i\Delta\alpha} \,,
\end{equation} 
where the phase shift
\begin{equation}
 \Delta\alpha = \frac{1}{\hbar} \int_i^f d\Lambda
\end{equation}
\end{subequations}
is independent of the path taken in the configuration space.
The indices $i$ and $f$ represent the initial and final points 
in the configuration space of the system, respectively.
Eq.~\eqref{eq:L_transform} is valid in general and is not limited to
the particular problem of the monopoles considered here.
A typical example is the gauge symmetry where the gauge field $A_\mu$ is
given by $A_\mu=\partial_\mu\Lambda$.
Eq.~\eqref{eq:L_transform} is also useful for deriving
the exchange phase factor of the charge-monopole composites.

The Lagrangian of two identical charge-monopole composites 
can be written as
\begin{equation}
 {\cal L} = {\cal L}_1 + {\cal L}_2 + {\cal L}_\mathrm{int}\, .
\label{eq:two_composites}
\end{equation}
For Type I, ${\cal L}_i$~($i=1,2$) represents each composite described by
the Lagrangian of Eq.~\eqref{eq:L_psi}. 
The inter-cluster interaction is given by 
\begin{equation}
 {\cal L}_\mathrm{int} 
  = (\dot{\vec{\psi}}_1 - \dot{\vec{\psi}}_2) \cdot 
  \left[ \mathbf{S}(\mathbf{r}) - \mathbf{S}(-\mathbf{r}) \right] \,, 
\end{equation}
where
$\dot{\vec{\psi}}_i$~($i=1,2$) is the angular velocity of each particle, and 
$\mathbf{S}(\mathbf{r}) = -(eg/c)\hat{\mathbf{r}}$ 
is the field angular momentum 
produced by the inter-cluster interaction.
For an exchange of two particles, 
$(\dot{\vec{\psi}}_1 - \dot{\vec{\psi}}_2) \cdot \hat{\mathbf{r}} = 0$,
and thus
${\cal L}_\mathrm{int}=0$: 
The system can be regarded as two independent composite particles.
Each particle contains the intrinsic
spin $s$ and field angular momentum of $n/2$ in units of $\hbar$.
The field angular momentum is also represented by 
a spinor~(see Eqs.~\eqref{eq:H} and \eqref{eq:H_nonrel}) 
in the Type-I charge-monopole cluster.
Therefore, the wave function has the symmetry
of Eq.~\eqref{eq:exchange}.

Derivation of the exchange symmetry for the Type-II system 
is equivalent to that presented in Ref.~\onlinecite{goldhaber76} 
analyzed with the vector potential $\mathbf{A}$. 
The Lagrangian of two identical composite particles
is also given in the form of Eq.~\eqref{eq:two_composites}. In our
formulation,
each term ${\cal L}_i$~($i=1,2$) for the composite particle 
involving the Type-II monopole is described
by the field-interaction Lagrangian \eqref{eq:L_Pi}. We find that
the inter-cluster interaction is 
\begin{equation}
 {\cal L}_\mathrm{int} = \dot{\mathbf{r}} \cdot 
 \left[ \mathbf{\Pi}(\mathbf{r}) - \mathbf{\Pi}(-\mathbf{r}) \right] \,,
\end{equation}
where $\mathbf{\Pi}(\mathbf{r})$($\mathbf{\Pi}(-\mathbf{r}))$ 
is the field momentum produced by
the charge at $\mathbf{r}_1$($\mathbf{r}_2$) interacting with 
the Type-II monopole at $\mathbf{r}_2$($\mathbf{r}_1$).
(Note that the two field momenta can be set identical without
affecting the result).
It is straightforward to show that 
$\mathbf{\Pi}(\mathbf{r})-\mathbf{\Pi}(-\mathbf{r})$ is curl-free:
$\mathbf{\Pi}(\mathbf{r})-\mathbf{\Pi}(-\mathbf{r}) = \nabla\Lambda$,
and thus ${\cal L}_\mathrm{int} = d\Lambda/dt$. 
Therefore, the wave function of the system is given in the form
\begin{equation}
 \Psi(\mathbf{r},\eta_1,\eta_2) = e^{i\alpha} \Psi_0(\mathbf{r},\eta_1,\eta_2)
 \,.
\label{eq:Psi-Psi_0}
\end{equation}
Unlike the Type-I system, the wave function $\Psi_0$ associated with
${\cal L}_1+{\cal L}_2$ does not include the field angular momentum in its
internal state, and satisfies
\begin{equation}
 P\Psi_0(\mathbf{r},\eta_1,\eta_2) 
  = (-1)^{2s} \Psi_0(\mathbf{r},\eta_1,\eta_2) \,.
\end{equation}
Applying Eq.~\eqref{eq:L_transform}, we obtain the interaction-induced
phase shift, 
\begin{eqnarray}
 \Delta\alpha &=&
   \frac{1}{\hbar} \int_C  
   \left[ \mathbf{\Pi}(\mathbf{r}) - \mathbf{\Pi}(\mathbf{-r}) \right]
   \cdot d\mathbf{r} \nonumber \\ 
   &=& \frac{1}{\hbar} \oint \mathbf{\Pi}(\mathbf{r})\cdot d\mathbf{r}
   = n\pi \,,
\end{eqnarray} 
where $C$ is a path for the exchange,
and therefore the wave function of the system has the symmetry of 
Eq.~\eqref{eq:exchange}.
In summary, the exchange symmetry of the composite particle 
satisfies the relation \eqref{eq:exchange} for both
Type-I and Type-II monopoles. An interesting feature is that
the phase factor $(-1)^{2s+n}$ is established in different ways for the two
types.

\subsection{Analogy to the Coriolis interaction}
Finally, we point out the formal equivalence between the
Type-I monopole and the Coriolis interaction.
This equivalence provides useful insight
into the physics of magnetic monopoles and their analogues.
Take an object of mass $m$ at point $P$ in a uniformly rotating frame 
with angular velocity $\vec{\omega}_0$~(Fig.~5). 
In the nonrelativistic limit, the Lagrangian of the object 
can be written as~(see e.g., Ref.~\onlinecite{landau60})
\begin{subequations}
\begin{equation}
 {\cal L} = \frac{1}{2}m \dot{\mathbf{r}}\cdot\dot{\mathbf{r}}
          + \frac{1}{2}m|\vec{\omega}_0\times\mathbf{r}|^2
          + {\cal L}_c \,,
\end{equation}
where the ``Coriolis" term ${\cal L}_c$ is 
\begin{equation}
 {\cal L}_c = m \dot{\mathbf{r}} \cdot \vec{\omega}_0\times\mathbf{r}\,.
\label{eq:L_c}
\end{equation}
The position $\mathbf{r}$ of the particle is specified by the 
locally flat coordinates with the basis vectors $\vec{e}_1$ and $\vec{e}_2$
(see Fig.~5):
$\mathbf{r} = u\vec{e}_1 + v \vec{e}_2$.
The angle vector $\vec{\psi}$ with $|\vec{\psi}|=\arctan{(v/u)}$
is perpendicular to $\mathbf{r}$,
and the Coriolis interaction of Eq.~\eqref{eq:L_c}
can be rewritten as
\begin{equation}
 {\cal L}_c = \dot{\vec{\psi}} \cdot \mathbf{L}_0 \,,
\label{eq:L_c_psi}
\end{equation} 
\end{subequations}
where $\mathbf{L}_0 = mr^2\vec{\omega}_0$ corresponds to the angular momentum
due to the rotation of the Lab frame with angular velocity $\vec{\omega}_0$.
This form of the Coriolis Lagrangian is equivalent to the
interaction Lagrangian of a Type-I monopole~(Eq.~\eqref{eq:L_psi}). 
The field angular momentum $\mathbf{S}$ in the charge-monopole pair
plays the same role as the mechanical angular momentum $\mathbf{L}_0$
in the Coriolis interaction (Eq.~\eqref{eq:L_c_psi}).

In the Coriolis interaction, a geometric phase shift already appears 
at the classical level: the precession angle $\Omega-2\pi$ 
($\Omega$ is the solid angle) of a Foucault
pendulum for one rotation of the frame. An interesting coincidence is found
between this precession angle and 
$\varphi$ in Eq.~\eqref{eq:varphi} for $S=\hbar$ ($n=2$).

\section{Conclusion} 
Adopting the local field interaction approach, we showed that the
Dirac monopole can be classified into two different types. 
Notably, the difference is not merely a mathematical artifact but a physical
reality, which arises from the presence or absence of a Dirac
string. The duality of the two types was analyzed, and it
revealed that
the two types yield identical results except a quantum phase factor $(-1)^n$ 
for one loop rotation depending on the quantum number $n$ of the field
angular momentum.
We also pointed out the formal equivalence of the Type-I monopole
with the Coriolis interaction.
For Type II, the Maxwell duality gives rise to various possibilities of 
the Dirac string configuration: 
Both the electric and magnetic charges may emerge as endpoints of the
corresponding strings.
The exchange symmetry was analyzed for identical particles of
charge-monopole composites. In both types, the wave function has a symmetry
factor of $(-1)^{2s+n}$, including the intrinsic spin ($s$) and the field
angular momentum.
Notably, our approach provides a physically realistic description of
the interaction between a charge and magnetic a monopole. This is possible
owing to the locality of the field-interaction theory~\cite{kang15,kang19}.


 \bibliography{references}

\begin{figure}
\centering
\includegraphics[width=8cm]{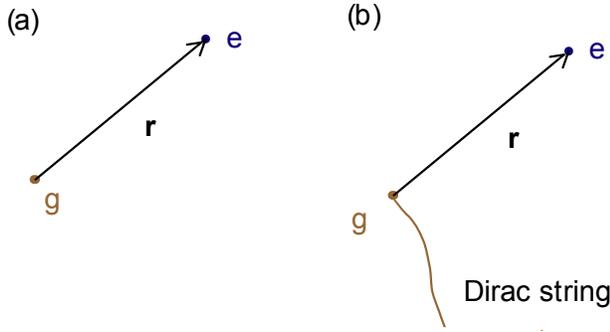} 
\caption{Electric charge $e$ and 
 a magnetic monopole $g$ (a) without singularity (``Type I"), and 
 (b) with a Dirac string attached to the latter (``Type II").
 }
\end{figure}
\begin{figure}
\centering
\includegraphics[width=6cm]{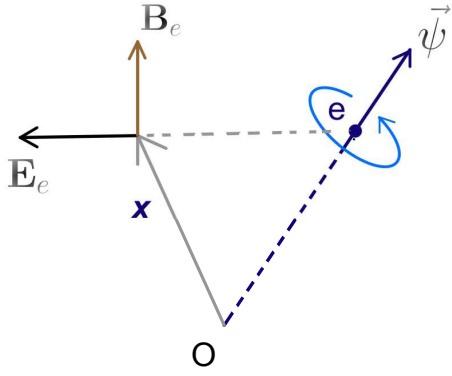} 
\caption{Illustration of a rotating (spinning) charge with
 angular velocity $\dot{\vec{\psi}}$. 
 Electric~($\mathbf{E}_e$) and magnetic~($\mathbf{B}_e$) fields
 are generated by the rotating charge at an arbitrary position $\mathbf{x}$.
 }
\end{figure}
\begin{figure}
\centering
\includegraphics[width=8cm]{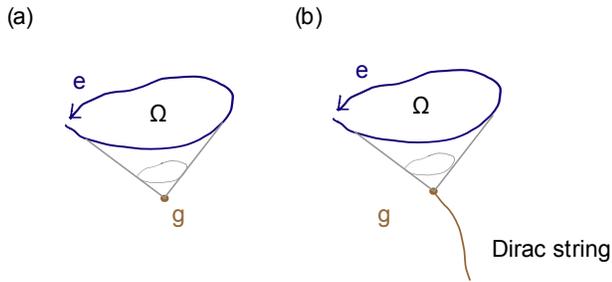} 
\caption{Arbitrary closed paths of the charge under (a) Type-I 
and (b) Type-II monopoles, respectively.
 }
\end{figure}
\begin{figure}
\centering
\includegraphics[width=8cm]{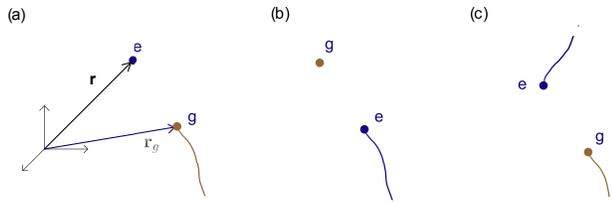} 
\caption{Various possible configurations of Dirac strings for describing
a Type-II system: 
(a) Magnetic monopole as an endpoint of
the invisible magnetic flux tube. (b) An electric charge emerges as
an endpoint of the invisible electric string. 
(c) Both particles are the endpoints of each string.
 All observable phenomena are independent of the string configuration.
 }
\end{figure}
\begin{figure}
\centering
\includegraphics[width=8cm]{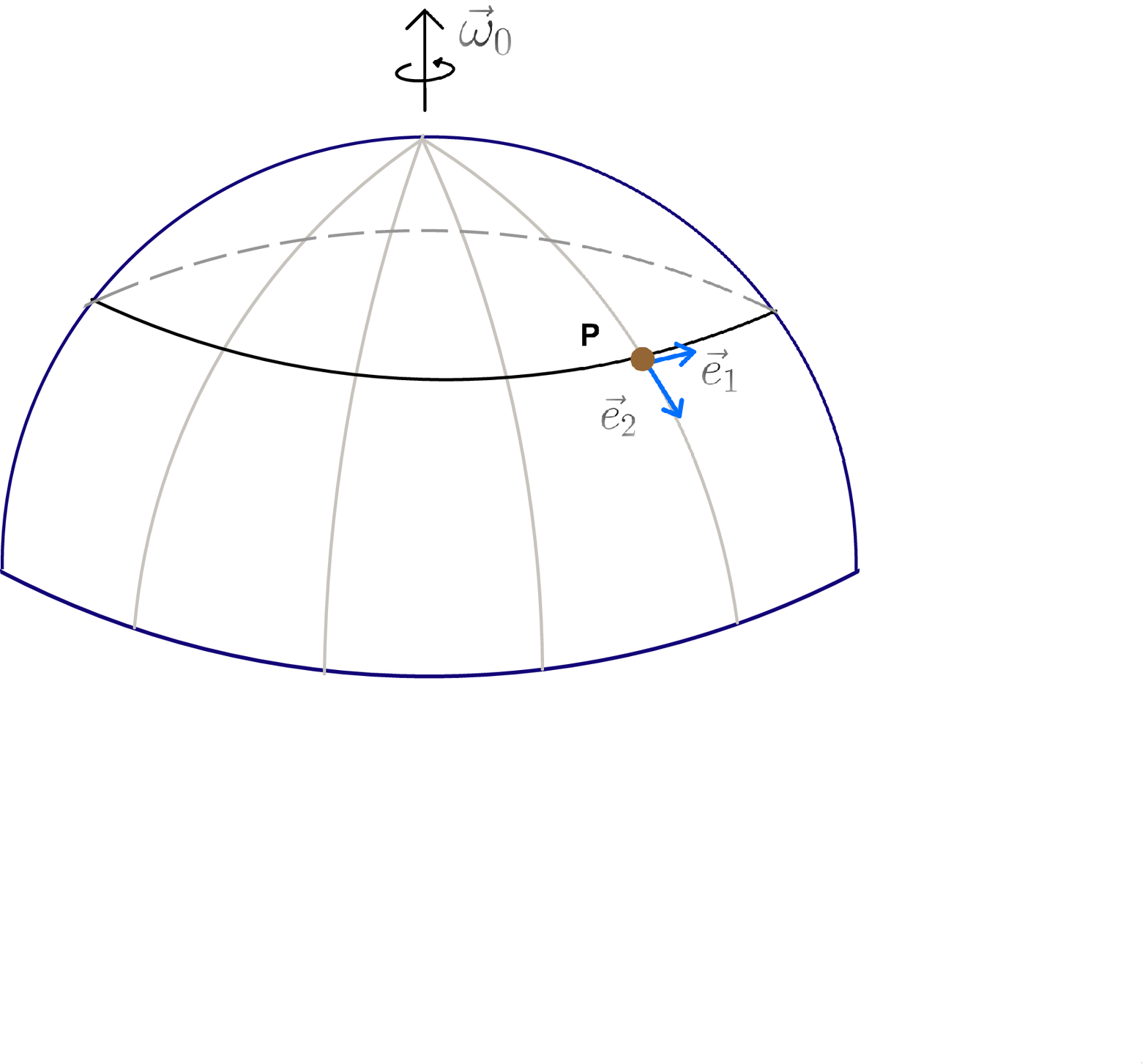} 
\caption{Description of the Coriolis interaction: A particle with mass $m$ under
a rotating frame (with angular velocity $\vec{\omega}_0$)
is described by locally flat coordinates: 
$\mathbf{r} = u\vec{e}_1 + v\vec{e}_2$.
 }
\end{figure}
\end{document}